\documentclass[AMA,STIX2COL,Linenumbersoff]{MRM}
\articletype{Technical Note}%

\usepackage{amsmath}
\usepackage[rightcaption]{sidecap}
\usepackage{amssymb}
\usepackage{float}
\usepackage{makecell}

\long\def\red#1{\bgroup \color{red}#1\egroup}

\def\todo#1{\bgroup \red{(TODO: #1)} \egroup}
\def\tocite{\bgroup \red{[TODO: ADD REF]}\egroup}

\begin{document}

\title{Looping Kooshball: Improved k-space coverage for silent fMRI using 3D coherence trajectories}

\author[1]{David J. Frey}{\orcid{0009-0001-8187-5962}}

\author[2]{Shreya Ramachandran}{\orcid{0000-0001-7023-1718}}

\author[1,3]{Jeffrey A. Fessler}{\orcid{0000-0001-9998-3315}}

\author[1,4]{Douglas C. Noll}{\orcid{0000-0002-0983-3805}}

\author[5]{Tobias C. Wood}{\orcid{0000-0001-7640-5520}}

\authormark{Frey, D.J. \textsc{et al}}

\address[1]{\orgdiv{Department of Biomedical Engineering}, \orgname{University of Michigan}, \orgaddress{\state{Ann Arbor, MI}, \country{United States}}}

\address[2]{\orgdiv{Department of Electrical Engineering and Computer Sciences}, \orgname{University of California, Berkeley}, \orgaddress{\state{Berkeley, California}, \country{United States}}}

\address[3]{\orgdiv{Department of Electrical Engineering and Computer Science}, \orgname{University of Michigan}, \orgaddress{\state{Ann Arbor, MI}, \country{United States}}}

\address[4]{\orgdiv{Department of Radiology}, \orgname{Michigan Medicine}, \orgaddress{\state{Ann Arbor, MI}, \country{United States}}}

\address[5]{\orgdiv{Department of Neuroimaging}, \orgname{King's College London}, \orgaddress{\state{London}, \country{United Kingdom}}}

\corres{David Frey, \email{djfrey@umich.edu}}

\finfo{This work was supported in part by \fundingAgency{National Institutes of Health} grant \fundingNumber{NIH-1R01EB035618}.}

\abstract[Summary]{
\section{Purpose}
To improve image quality in near-silent fMRI over conventional Looping Star (LS) by using a periodic 3D Looping Kooshball (LKB) k-space encoding trajectory.

\section{Methods}
Conventional Looping Star (LS), comprised of rotated 2D planar loops, and Looping Kooshball (LKB) trajectories, comprised of rotated 3D loops, were compared using point-spread-function (PSF) analysis, phantom imaging, acoustic measurements, and visual-motor fMRI in three participants.

\section{Results}
At 3-mm isotropic resolution and a 1-s volume TR, LKB produced a more
spatially concentrated point-spread function, reduced phantom NRMSE by
18.5\%, and increased SSIM and HaarPSI~\cite{reisenhofer2018haar} by 2.0\% and 12.2\%, respectively.
In vivo, LKB produced a mean within-participant increase of 29.4\% in median tSNR and increased mean ROC AUC from 0.853 to 0.907; both metrics improved in three of four participants.
Acoustic levels ($L_\mathrm{Aeq}$) were 66.0 dBA for LS, 73.4 dBA for LKB, and 64.8 dBA under ambient conditions.

\section{Conclusion}
LKB improved volumetric sampling and image quality relative to conventional LS, with preliminary evidence of improved fMRI performance at the cost of modestly increased acoustic noise.
}


\wordcount{2800}

\maketitle

\section{Introduction}\label{sec1}

Functional MRI (fMRI) has become a cornerstone of neuroscience, enabling researchers to characterize brain activity and connectivity in health and disease without invasive surgical intervention~\cite{du2015group, oldehinkel2013resting, zhan2015window, wiebking2015neural, hacker2012resting}. However, the inherent tradeoffs between spatial resolution, temporal resolution, signal-to-noise ratio (SNR), and hardware constraints limit what can be achieved in practice.

Gradient echo-echo planar imaging (GE-EPI)-based pulse sequences, specifically Simultaneous Multi-slice-EPI (SMS-EPI), have become the standard pulse sequences for fMRI due to their high readout efficiency and sensitivity to T2*-weighted BOLD contrast~\cite{moeller2010multiband, feinberg2012rapid, setsompop2012blipped, iraji2015resting}.
However, GE-EPI achieves this efficiency in part by driving gradient hardware to its operating limits, producing acoustic sound pressure levels capable of damaging human hearing~\cite{mcjury2000auditory}.
Beyond the risk of hearing damage, scanner noise induces patient anxiety~\cite{nguyen2020prevalence,mcnulty2009acoustic}, particularly in noise-sensitive populations~\cite{golden2015auditory, hamann2002impaired,stogiannos2023toward,reeve2025anxiety,liberzon1999brain,chen2021neuroimaging}.
Even in healthy volunteers, noise-related anxiety increases the likelihood of motion during the scan, introducing artifacts that can compromise research findings or clinical diagnoses~\cite{d2026systematic}.

Critically, the presence of loud acoustic noise violates the assumption of task isolation underlying fMRI analysis; i.e., the assumption that measured BOLD response reflects only the intended stimulus, while resting-state studies assume the absence of structured external stimulation entirely.
Loud acoustic noise drives secondary neural processing in auditory pathways, suppresses primary task responses, and introduces non-localized confounding activations.
These effects have been demonstrated even in non-auditory paradigms, where scanner noise reduces the detectability of true BOLD activation and increases measurement variance~\cite{ke2021effect,brungart2001informational}.

Several approaches have been proposed to mitigate scanner noise. Physical hearing protection (earplugs, headphones, active noise cancellation~\cite{mansfield1994active,almeer2022mri}) reduces the risk of hearing damage but does not eliminate neural responses to acoustic stimulation or completely address patient anxiety. Reducing gradient slew rates can substantially lower acoustic noise~\cite{edelstein2002making,heismann2015sequence}, although the strongest reductions often come at the expense of temporal resolution. Using sinusoid waveforms instead of trapezoids can improve the acoustic characteristics of EPI, but such sequences typically remain considerably louder than ambient conditions~\cite{hutter2018quiet}.

Looping Star (LS) is a T2*-weighted pulse sequence designed for near-silent fMRI acquisition\cite{wiesinger2019looping,damestani2021coherence,wiesinger2023book}.
It is well suited to noise-sensitive populations and has been shown to reduce inadvertent contamination from auditory pathway activation\cite{dionisio2020looping,damestani2021revealing}.
LS achieves favorable spatiotemporal resolution by multiplexing temporally staggered coherences encoded with slowly varying gradient waveforms, thereby limiting acoustic noise.
Conventional LS uses a planar encoding trajectory for each shot.
Because the multiplexing scheme precludes conventional slice-selective excitation, the encoding plane is rotated across shots to obtain 3D coverage.
All rotated planes intersect near the k-space origin, resulting in dense sampling of low spatial frequencies but relatively sparse and nonuniform sampling of peripheral 3D k-space.
This limits sampling efficiency, particularly when few shots are combined to achieve high temporal resolution.
Previous work explored Wave Looping Star, which added through-plane modulation to broaden each nominal sampling plane and improve volumetric coverage~\cite{solana2024silent}.

In this work, we explore a complementary approach termed Looping Kooshball (LKB), in which the encoding trajectory extends through all three spatial dimensions within each shot.
LKB is designed to improve volumetric k-space coverage while preserving the echo-forming and coherence-crushing properties of conventional LS.
We compare LKB with conventional LS using point-spread-function analysis, in-bore acoustic measurements, structural phantom imaging, and preliminary visual-motor fMRI.
We hypothesized that LKB would reduce undersampling artifacts and improve temporal stability and activation detection while retaining the favorable acoustic characteristics of LS.

\section{Theory}\label{sec2}

\begin{figure*}[t]
    \centering
    \includegraphics[]{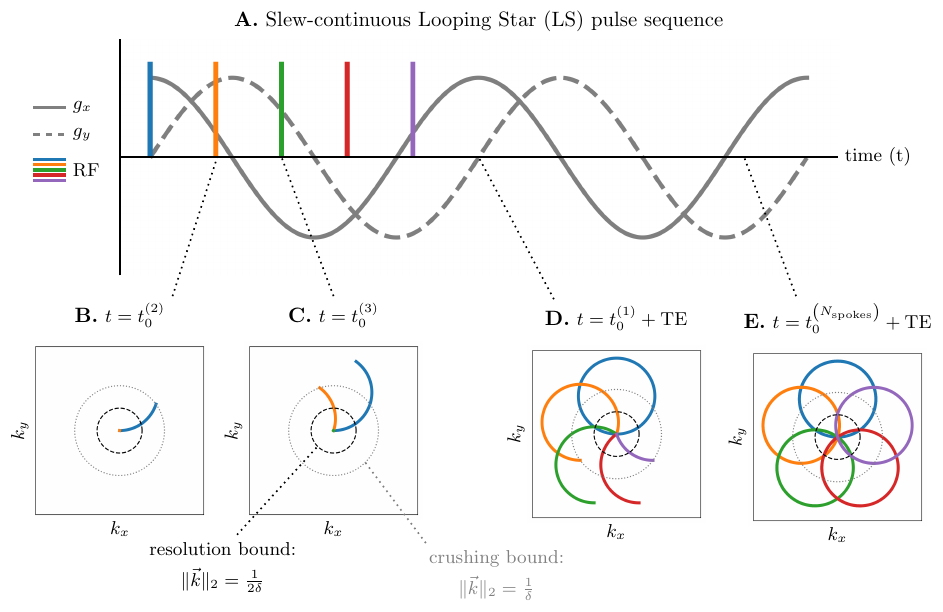}
    \caption{Conventional slew-continuous Looping Star (LS) pulse sequence and k-space-trajectory evolution. \\
    \textbf{A.} Encoding gradients and RF pulses for one shot of single-echo LS.
    Colors identify the different excited coherences. \\
    \textbf{B.} Trajectories at the excitation of the second coherence, $t=t_0^{(2)}$.
    The first coherence has reached the crushing threshold, $\|\vec{k}\|_2=1/\delta$, and is therefore assumed to be sufficiently dephased to suppress interference from the second excitation. \\
    \textbf{C.} Trajectories at the excitation of the third coherence, $t=t_0^{(3)}$.
    The second coherence has reached the crushing threshold and encodes a direction distinct from the first. \\
    \textbf{D.} Trajectories at the first echo, $t=t_0^{(1)}+\mathrm{TE}$.
    The first coherence has returned to the k-space origin, whereas the other coherences remain outside the crushing threshold, reducing contamination of its acquired signal. \\
    \textbf{E.} Trajectories at the last echo, $t=t_0^{(N_\mathrm{spokes})}+\mathrm{TE}$.
    All coherences have been refocused at their respective echo times, producing a star-shaped k-space sampling pattern.}
    \label{fig:fig1}
\end{figure*}

As illustrated in Figure~\ref{fig:fig1},
the conventional slew-continuous Looping Star (LS) pulse sequence~\cite{wiesinger2019looping,dionisio2020looping} uses a burst of several RF pulses to excite $N_\mathrm{spokes}$ signal coherences, equally spaced at times $t_0^{(s)} = (s{-}1)\,\mathrm{TE}/N_\mathrm{spokes},\ \forall s = 1,\ldots N_\mathrm{spokes}$, given echo time, $\mathrm{TE}$.
Oscillating gradients are played to refocus each coherence at its respective echo time, $t_0^{(s)} + \mathrm{TE},\ \forall s$.
The k-space trajectory of each spoke $s$ is defined in terms of the encoding trajectory, $\vec{k}_\mathrm{enc}(t): \mathbb{R} \to \mathbb{R}^3$:
\begin{equation} \label{eq:general_traj}
    \vec{k}^{(s)}(t) = \vec{k}_\mathrm{enc}(t) - \vec{k}_\mathrm{enc}\left(t_0^{(s)}\right), \quad t > t_0^{(s)}.
\end{equation}

The conventional LS trajectory encodes a circle around the k-space origin:
\begin{equation} \label{eq:ls_enc}
    \vec{k}_{\mathrm{enc},\text{LS}}(t) = k_\mathrm{scale,LS} \begin{bmatrix}
            \sin \left( 2 \pi \frac{t}{\mathrm{TE}} \right) \\
            -\cos \left( 2 \pi \frac{t}{\mathrm{TE}} \right) \\
            0
        \end{bmatrix},
\end{equation}
where $k_\mathrm{scale} \in \mathbb{R}$ is a constant scaling term.

Two properties are central to the design of the LS encoding.
First, each spoke's k-space trajectory returns
to the origin at integer multiples of $\mathrm{TE}$ after excitation, forming a gradient echo (GRE):
\begin{equation} \label{eq:echo_condition}
    \begin{aligned}    
        \vec{k}^{(s)}&\left(t_0^{(s)} + n\mathrm{TE}\right) = \vec{0}, \\
        &\forall n = 0,\ldots N_\mathrm{echoes}, \\
        &\forall s = 1,\ldots,N_\mathrm{spokes}.
    \end{aligned}
\end{equation}
Echoes are indexed by $n = 0, \ldots, N_\mathrm{echoes}$, where $n=0$ corresponds to the initial free induction decay (FID) signal and $n = 1, \ldots, N_\mathrm{echoes}$ correspond to the GRE signals.

Second, each coherence accumulates sufficient dephasing between excitations to suppress interference from the other excited coherences.
We empirically define a coherence as sufficiently crushed when it accumulates at least one full cycle of phase variation across a nominal resolution element of width $\delta$, measured along the direction of its k-space displacement.
Therefore, we impose the empirical crushing constraint
\begin{equation}
\label{eq:crush_condition}
\begin{aligned}
    \|\vec{k}^{(s)}(t)\|_2
    \geq \frac{1}{\delta},
    \qquad
    &\forall t \notin
    \bigcup_{n=0}^{N_\mathrm{echoes}}\mathcal{T}^{(s,n)}, \\
    &\forall s=1,\ldots,N_\mathrm{spokes},
\end{aligned}
\end{equation}
where $\mathcal{T}^{(s,n)}$ denotes the active readout window for coherence $s$ at echo $n$.
The condition can be satisfied by solving for the appropriate closed form of $k_\mathrm{scale}$ such that the minimum distance between coherences in k-space is greater than or equal to the crushing bound $\frac{1}{\delta}$:
\begin{equation}
    k_\mathrm{scale} = \frac{\left(1/\delta \right)}{\underset{s \in \{1,\ldots,N_\mathrm{spokes}\}}{\min} \left\|\vec{k}^{(s)}\left(t_0^{(\operatorname{mod}(s,N_\mathrm{spokes})+1)}\right)\right\|_2}.
\end{equation}

To encode 3D k-space, the gradient channels are rotated for each shot according to a 3D golden-angle
schedule~\cite{feng2022golden} and binned together to form spherical sampling volumes. 
Because all spokes lie in rotated planes, the binned 3D k-space trajectory densely samples the origin where all planes intersect, but poorly samples the edges of the sphere resulting in the contamination of high spatial frequency aliasing artifacts in each image volume.

To improve 3D k-space sampling, we propose an extended encoding trajectory termed Looping Kooshball (LKB).
The name reflects its out-of-plane sampling pattern which resembles kooshball trajectories in 3D radial MRI literature~\cite{feng2022golden}.
The LKB encoding trajectory can be expressed as
\begin{equation} \label{eq:kooshball}
    \begin{aligned}
        &\vec{k}_{\mathrm{enc},\text{LKB}}(t) \\
        = &k_\mathrm{scale,LKB} \begin{bmatrix}
                \frac{1}{3}\sin \left( 2 \pi \frac{t}{\mathrm{TE}} \right) + \sin \left( 6 \pi \frac{t}{\mathrm{TE}} \right) \\
                -\frac{1}{3}\cos \left( 2 \pi \frac{t}{\mathrm{TE}} \right) + \cos \left( 6 \pi \frac{t}{\mathrm{TE}} \right)\\
                \sin \left( 4 \pi \frac{t}{\mathrm{TE}} \right)
            \end{bmatrix}.
        \end{aligned}
\end{equation}
Because this trajectory is constructed from sinusoidal components with integer harmonics of $\frac{1}{\mathrm{TE}}$, it is also periodic with period $\mathrm{TE}$, satisfying the echo-forming condition in Equation~\eqref{eq:echo_condition} by construction.

\begin{figure*}[t]
    \centering
    \includegraphics[]{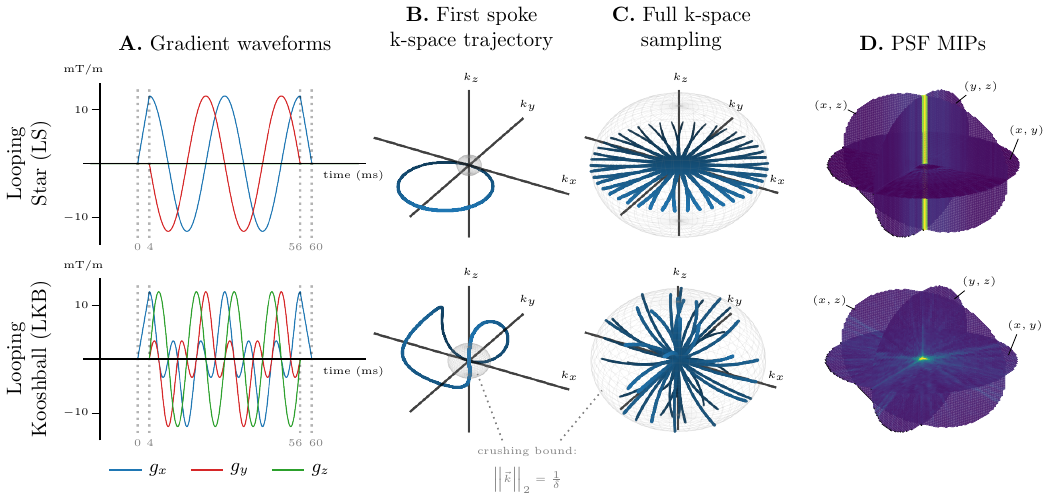}
    \caption{Conventional Looping Star (LS, top row) vs. Looping Kooshball (LKB, bottom row). \\
    Column \textbf{A.} Gradient waveforms with appended linear ramps. \\
    Column \textbf{B.} k-space trajectory of first spoke (coherence), i.e., $\vec{k}^{(1)}(t),\ \forall t \in \left[t_0^{(1)},t_0^{(1)}+\mathrm{TE}\right)$. \\
    Column \textbf{C.} Full set of k-space sampling coordinates inside the crushing bound, i.e., $\bigcup_s \left\{ \vec{k}^{(s)}(t),\ \forall t: \|\vec{k}^{(s)}(t)\|_2 \leq \frac{1}{\delta}\right\}$. \\
    Column \textbf{D.} Maximum intensity projections (MIPs) for the point spread function (PSF), reconstructed from a single rotation of the encoding trajectory.
    }
    \label{fig:fig2}
\end{figure*}

Figure~\ref{fig:fig2} shows
the gradient waveforms, k-space trajectory and sampling for both Looping Star variants (LS, LKB) alongside their point spread functions.
For both cases, linear ramps are appended to the beginning and end of the gradients to start and end at zero.

\section{Methods}

\subsection{Acquisition and reconstruction}

Both tested Looping Star trajectory variants (LS, LKB) were designed with $N_\mathrm{spokes}=23$, $N_\mathrm{echoes}=1$, echo time $\mathrm{TE}=26$ ms, nominal isotropic resolution $\delta=3$ mm, and gradient raster time $\tau=4~\mu$s.

Imaging data was acquired on a GE MR750 3T system with a Nova 32-channel head coil at the University of Michigan, Ann Arbor.
Acoustic measurements were performed on a GE MR750w 3T system at the University of California, Berkeley.

Images were reconstructed using the conjugate-gradient (CG) algorithm to solve the Tikhonov-regularized least-squares problem:
\begin{equation}
\label{eq:recon}
    \mathbf{x}_*
    =
    \underset{\mathbf{x}\in\mathbb{C}^{N\times V}}
    {\operatorname{argmin}}
    \;
    \frac{1}{2}
    \left\|\mathcal{A}(\mathbf{x})-\mathbf{b}\right\|_F^2
    +
    \frac{\beta}{2}
    \left\|\mathbf{x}\right\|_F^2,
\end{equation}
where the columns of $\mathbf{x}\in\mathbb{C}^{N\times V}$ contain the $N$-voxel images for $V$ reconstructed volumes, and $\mathbf{b}\in\mathbb{C}^{MQ\times V}$ contains the corresponding complex-valued k-space measurements from $Q$ receiver channels at $M$ sampling locations per volume.
The forward model $\mathcal{A}:\mathbb{C}^{N\times V}\rightarrow
\mathbb{C}^{MQ\times V}$ independently applies a sensitivity-weighted nonuniform fast Fourier transform (NUFFT) to each volume and stacks the resulting measurements over receiver channel dimension~\cite{fessler2007nufft,pruessmann1999sense}.
The scalar $\beta\geq0$ is the Tikhonov regularization parameter.

\subsection{Point-spread-function analysis}

Effective point-spread functions (PSFs) were estimated for both trajectory variants using the regularized reconstruction in Equation~\eqref{eq:recon}.
To isolate differences arising from the sampling trajectories, the PSF analysis used a single-coil NUFFT forward model with
uniform coil sensitivity.
The synthetic k-space data were set to $\mathbf{b}=\mathbf{1}$, corresponding to the Fourier samples generated by a unit impulse at the center of image space.

PSFs were reconstructed on an $N=80\times80\times80$ ($N=80^3$) isotropic grid with 3-mm voxel size.
Single-volume PSFs were reconstructed for several bin sizes (trajectory rotations per volume) to evaluate the effect of increased temporal resolution and corresponding k-space undersampling on PSF quality.

PSF quality was quantified using a generalized $\ell_p$ integrated sidelobe ratio~\cite{lukin2012integrated}  ($\ell_p$-ISLR):
\begin{equation} \label{eq:islr}
    \ell_p\text{-ISLR} = \frac{\left\|(\mathbf{1}_N-\mathbf{w})\odot\boldsymbol{\Psi}\right\|_p}{\left\|\mathbf{w}\odot\boldsymbol{\Psi}\right\|_p},
\end{equation}
where $\boldsymbol{\Psi}\in\mathbb{C}^{N}$ is the reconstructed PSF, $\mathbf{1}_N\in\mathbb{R}^{N}$ is a vector of ones, $\mathbf{w}\in[0,1]^N$ is a center-lobe weighting vector,
and $\odot$ denotes element-wise vector multiplication.
We empirically defined the weighting vector by a 3D isotropic Gaussian function centered at the origin with $\sigma=1.5\text{ voxels}$.
We empirically selected $p=6$ to emphasize high-amplitude sidelobes while retaining sensitivity to spatially distributed sidelobe energy.
Lower $\ell_6$-ISLR values indicate greater concentration of the reconstructed response near the nominal point-source location.

\subsection{Structural phantom experiments}

A small ACR MRI phantom~\cite{ihalainen2011mri} was scanned using both trajectory variants.
Images were reconstructed according to Equation~\eqref{eq:recon} on a $54\times54\times54$ isotropic grid ($N=54^3$) with 3-mm voxel size, using $Q=12$ compressed virtual receiver channels, $\beta=2^{14}$, and 30 CG iterations.

Images were reconstructed using several bin sizes (trajectory rotations per volume) to evaluate the effect of increased temporal resolution and corresponding k-space undersampling on image quality.
Normalized root-mean-square error (NRMSE), structural similarity index (SSIM), and Haar perceptual similarity index (HaarPSI)~\cite{reisenhofer2018haar} were calculated relative to a reference image formed by averaging the 4800-rotation reconstructions from both trajectories.

\subsection{In-bore acoustic measurements}

An MR-compatible free-field condenser microphone (Model 2541; Larson Davis, Depew, NY) was positioned inside the scanner bore at the approximate location of the participant's right ear.
The microphone was connected by a 10-m cable to an integrating sound-level meter (Model 831C; Larson Davis) located in the console room.
Measurements were acquired during ambient conditions and during operation of both trajectory variants.
The A-weighted equivalent continuous sound pressure level ($L_{\mathrm{Aeq}}$) and A-weighted acoustic spectrum were measured over 30-s intervals.

\subsection{In vivo fMRI experiments}

Four healthy participants were scanned while performing a block-design visual-motor task consisting of a flashing checkerboard accompanied by bilateral finger tapping, alternating between 20-s task and 20-s rest blocks.
For both trajectory variants, two 5-min runs were acquired with a shot repetition time of 62.5 ms, yielding 4800 trajectory rotations per run.
The trajectory rotation schedule repeated every 20 s.

Images were reconstructed according to Equation~\eqref{eq:recon} on an $80\times80\times80$ isotropic grid ($N=80^3$) with 3-mm voxel size, using $Q=12$ compressed virtual receiver channels, $\beta=2^{14}$, and 30 CG iterations.

For both trajectory variants, the two 5-min image time series were concatenated and processed using UNFOLD spectral filtering to suppress periodic aliasing artifacts~\cite{frey2025improved,madore2004unfold}.
Voxelwise activation $t$ statistics were calculated using a general linear model (GLM) containing a constant and task regressor.
Temporal SNR (tSNR) was calculated from the post-UNFOLD time series after regression of the task signal.

A reference fMRI run was also acquired using a simultaneous multislice gradient-echo EPI (SMS-EPI) sequence with matched nominal spatial and temporal resolution ($\delta = 3$ mm, TR = 1s).
A reference-positive activation mask was generated by thresholding the co-registered SMS-EPI activation map. This mask was used to calculate median tSNR within the reference-positive regions and to define the positive class for receiver operating characteristic (ROC) analysis of both trajectory variants.

\section{Results}

\subsection{Point-spread-function analysis}

Figure~\ref{fig:fig3}B shows $\ell_6$-ISLR as a function of bin
size, defined as the number of k-space trajectory rotations per volume.
The LKB trajectory produced lower $\ell_6$-ISLR than conventional LS at every
evaluated bin size. At 16 rotations per volume, corresponding to a volume
TR of 1 s, $\ell_6$-ISLR decreased from 0.208 for LS to 0.180 for LKB, a relative decrease of 13.5\%.

\subsection{Structural phantom experiments}

\begin{figure*}[t]
    \centering
    \includegraphics[]{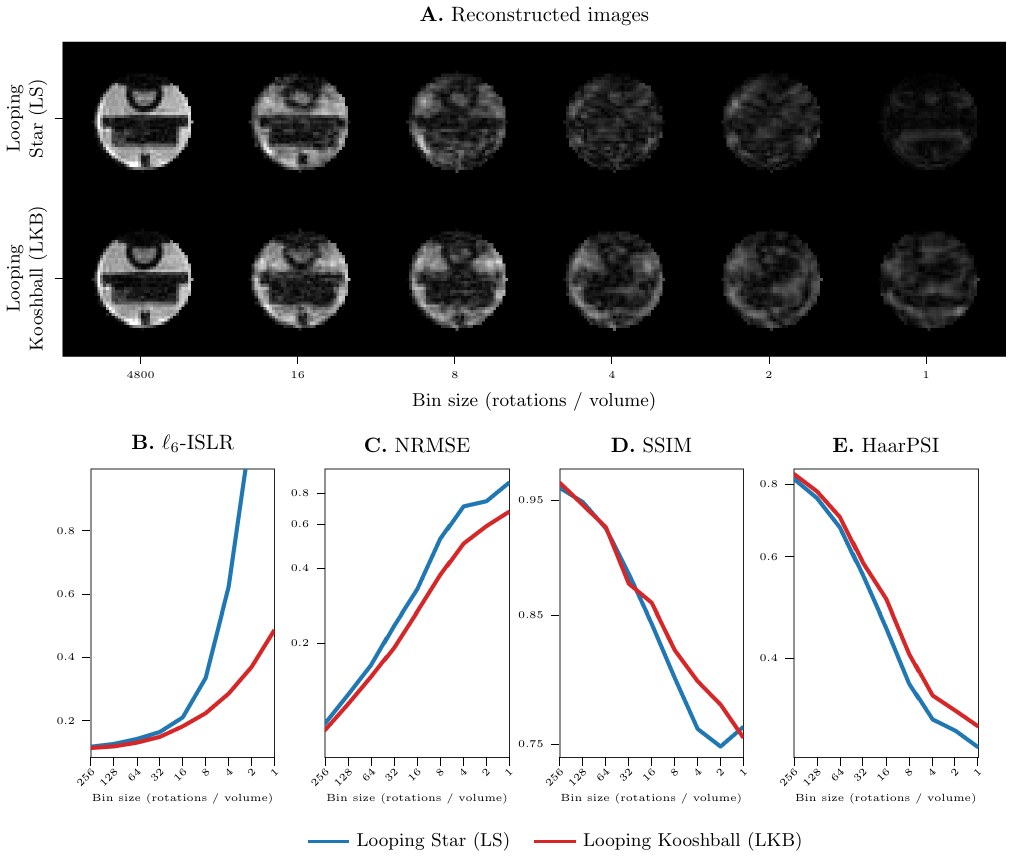}
    \caption{Point-spread-function and ACR phantom results as a function of
    bin size, defined as the number of k-space trajectory rotations
    contributing to each 3D volume. \\
    \textbf{A.} Representative axial slices from the phantom
    reconstructions. \\
    \textbf{B.} $\ell_6$-ISLR of the reconstructed PSFs. Lower values
    indicate greater concentration of the reconstructed response within the
    center-lobe region. \\
    \textbf{C--E.} NRMSE, SSIM, and HaarPSI, respectively, for the phantom
    reconstructions. Image-quality metrics were calculated relative to
    well-sampled reference reconstructions formed using all 4800
    rotations.}
    \label{fig:fig3}
\end{figure*}

Figure~\ref{fig:fig3} shows ACR phantom reconstructions as a function of the
number of trajectory rotations included in each image. At every evaluated
bin size, LKB produced lower NRMSE and higher HaarPSI than conventional LS. The
LKB trajectory also produced higher SSIM at most bin sizes, with the
largest differences observed at smaller bin sizes.

At 16 rotations per volume, corresponding to a volume TR of 1 s, NRMSE
decreased from 0.329 for LS to 0.268 for LKB, an 18.5\% relative
decrease. SSIM increased from 0.843 to 0.860, a 2.0\% relative increase,
and HaarPSI increased from 0.452 to 0.507, a 12.2\% relative increase.

Qualitatively, the LKB images exhibited less geometric distortion and fewer aliasing artifacts than the LS images, especially at smaller bin sizes.

\subsection{In-bore acoustic measurements}

\begin{figure*}[t]
    \centering
    \includegraphics[]{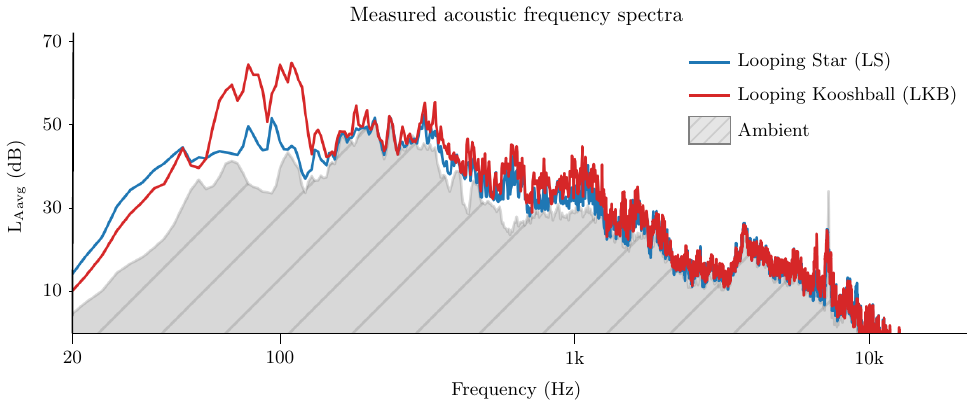}
    \caption{A-weighted acoustic spectra measured over 30-s intervals under
    ambient conditions and during operation of the LS and LKB
    trajectories.}
    \label{fig:fig4}
\end{figure*}

Figure~\ref{fig:fig4} shows the A-weighted acoustic spectra for ambient noise, LS, and LKB.
The measured
$\mathrm{L}_{\mathrm{Aeq}}$ values were 64.8 dBA for ambient noise,
66.0 dBA for LS, and 73.2 dBA for LKB.
The acoustic level of LKB was therefore 7.2 dB higher than that of LS and 8.4 dB above the
ambient level.
The largest spectral differences between the trajectory
variants occurred between 100 and 130 Hz, encompassing the frequency of the
highest harmonic component of the LKB encoding trajectory,
$3/\mathrm{TE}\approx115$ Hz.

\subsection{In vivo fMRI experiments}

\begin{figure*}[t]
    \centering
    \includegraphics[]{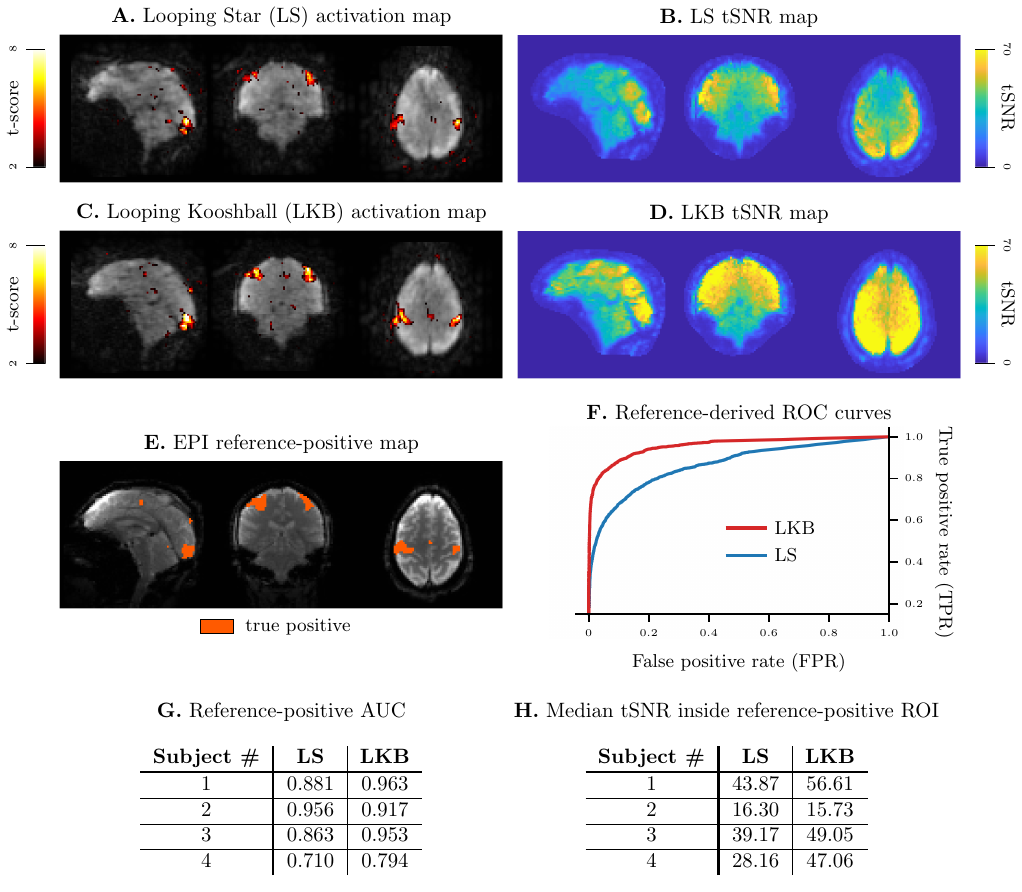}
    \caption{In vivo visual-motor task fMRI results for first participant only. Orthogonal slices are
    shown and include EPI-derived
    reference-positive regions in the primary visual and motor cortices. \\
    \textbf{A.} Activation $t$ statistics for LS overlaid on the first
    image volume before UNFOLD filtering. \\
    \textbf{B.} LS tSNR after UNFOLD filtering and regression of the task
    signal. \\
    \textbf{C,D.} Activation $t$ statistics and post-UNFOLD tSNR,
    respectively, for LKB, displayed using the same scales as in
    \textbf{A,B}. \\
    \textbf{E.} EPI-derived reference-positive activation mask overlaid on
    the first volume of the reference SMS-EPI acquisition. \\
    \textbf{F.} ROC curves for LS and LKB. The positive class was defined by the EPI-derived reference-positive mask shown in \textbf{E}. \\
    \textbf{G.} Participant level EPI-derived reference-positive AUC values. \\
    \textbf{H.} Participant level median tSNR values within the EPI-derived reference-positive ROIs. \\
    }
    \label{fig:fig5}
\end{figure*}

The LKB trajectory generally produced higher temporal stability and visual-motor activation-detection performance than conventional LS (Figure~\ref{fig:fig5}).

Figure~\ref{fig:fig5}G reports
participant level EPI-derived reference-positive AUC values.
Mean AUC increased from $0.853\pm0.103$ with LS to $0.907\pm0.078$ with LKB, with a mean paired increase of $0.054\pm0.062$. AUC increased with LKB in three of four participants.
The ROC curves for Participant 1 shown in Figure~\ref{fig:fig5}\textbf{F} illustrate activation-detection performance across thresholds.

Figure~\ref{fig:fig5}H reports
participant level median tSNR within the EPI-derived reference-positive regions.
Across participants, mean median tSNR increased from $31.87\pm12.30$ with LS to $42.09\pm18.04$ with LKB.
The mean paired increase was $10.21\pm8.12$, corresponding to a mean within-participant relative increase of 29.4\%.
tSNR increased with LKB in three of four participants.

Qualitatively, the LKB single-volume images exhibited fewer structured aliasing artifacts than the LS images.
The corresponding activation maps showed stronger visual-motor activation within the EPI-derived reference-positive regions and fewer extracortical clusters.

\section{Discussion}
We introduced a periodic 3D encoding trajectory called Looping Kooshball (LKB), to improve image quality in near-silent fMRI over conventional Looping Star (LS).
Unlike LS, which combines rotated planes that intersect near the k-space origin, LKB encodes all three spatial dimensions within each shot.
This reduces oversampling of central k-space and improves peripheral coverage, particularly when few rotations are available per volume.

The lower $\ell_6$-ISLR and improved phantom metrics obtained with LKB support this interpretation. Differences were generally greatest at
smaller bin sizes, indicating that 3D encoding is most beneficial under
greater undersampling. Preliminary in vivo results were consistent with the
PSF and phantom findings: mean participant-level median tSNR increased by
29.4\%, and mean ROC AUC increased from 0.853 to 0.907. Both metrics
improved in three of four participants.

Participant 2 exhibited substantially lower tSNR under both trajectories
and intermittent frame corruption across acquisitions. Both tSNR and AUC
decreased slightly with LKB in this participant, limiting confidence
in the within-participant comparison. Participant 2 was nevertheless
retained to avoid post hoc exclusion based on the observed outcomes.
Because only four participants and one visual-motor task were evaluated,
the in vivo findings should be interpreted as an initial feasibility
demonstration.

Improved sampling was accompanied by increased acoustic noise. The measured
$L_{\mathrm{Aeq}}$ was 73.2 dBA for LKB, 7.2 dB higher than for
conventional LS. Unlike conventional LS, LKB contains nonzero higher-order
harmonic components, including transverse components at
$3/\mathrm{TE}\approx115$ Hz. Differentiation of these components to obtain
the gradient waveforms amplifies their contribution to gradient slew rate
in proportion to frequency, shifting gradient spectral energy toward the
more acoustically sensitive frequency range. The largest measured acoustic
increase occurred near 115 Hz, consistent with this mechanism, although
the resulting sound level also depends on the scanner-specific mechanical
transfer function and gradient-coil resonances. LKB therefore presents a
tradeoff between improved sampling efficiency and increased acoustic noise,
whose suitability will depend on the application and scanner platform.

Equations~\eqref{eq:echo_condition} and~\eqref{eq:crush_condition} define two
general requirements for Looping Star-type trajectories: periodic refocusing at
the desired echo times and sufficient coherence separation outside active
readout windows. Both LS and LKB trajectories satisfy these requirements, suggesting a
broader family of compatible encoding 3D trajectories which are compatible with the Looping Star framework. Future designs could
jointly optimize sampling uniformity, coherence
separation, and acoustic response under the constraints in Equations~\eqref{eq:echo_condition} and~\eqref{eq:crush_condition}. Periodicity also permits extension to
the initial FID and subsequent gradient echoes.

Limitations include the small in vivo sample, use of SMS-EPI as an
imperfect activation reference, fixed reconstruction parameters, and
scanner dependence of the acoustic measurements. Moreover, acoustic and
imaging experiments were conducted on different systems. Larger studies
and scanner-specific trajectory optimization are warranted.

\section{Conclusions}
We introduced a periodic 3D encoding trajectory called Looping Kooshball (LKB), to improve image quality in near-silent fMRI over conventional Looping Star (LS).
At 3-mm isotropic resolution and a 1-s volume TR, LKB reduced effective PSF $\ell_6$-ISLR by 13.5\% and phantom NRMSE by 18.5\%, while increasing phantom SSIM and HaarPSI by 2.0\% and 12.2\%, respectively, relative to conventional LS.
In preliminary task-fMRI experiments, LKB produced a mean within-participant increase of 29.4\% in median tSNR and increased mean ROC AUC from 0.853 to 0.907, with both metrics improving in three of four participants.
These benefits were accompanied by a 7.2-dB increase in $L_{\mathrm{Aeq}}$, resulting in a measured acoustic level of 73.2 dBA.
Thus, LKB provides a tradeoff between improved spatiotemporal sampling efficiency and increased acoustic noise and motivates further optimization of this tradeoff.
These results establish its feasibility as an improved volumetric encoding strategy for near-silent fMRI and motivate evaluation in larger cohorts and additional functional paradigms.

\subsection*{Acknowledgments}
The authors would like to thank Scott J. Peltier for help acquiring early in-bore acoustic noise measurements, Yongli He for help acquiring reference EPI fMRI datasets, and Florian Wiesinger for insightful discussions on Looping Star.

\subsection*{Data availability statement}
A Pulseq~\cite{layton2017pulseq} implementation of the Looping Star pulse sequence, compatible with the Looping Kooshball trajectory, is available as a public repository at: \url{https://www.github.com/fmrifrey/lps}.

\subsection*{Conflicts of Interest}
The authors declare no conflicts of interest.

\bibliography{references}%

\end{document}